\documentclass[%
 aip,
 amsmath,amssymb,
 reprint,%
]{revtex4-1}

\usepackage{graphicx}
\usepackage{dcolumn}
\usepackage{bm}
\usepackage{booktabs}

\usepackage[utf8]{inputenc}
\usepackage[T1]{fontenc}
\usepackage{mathptmx}
\usepackage{etoolbox}
\usepackage{microtype}

\makeatletter
\def\@email#1#2{%
 \endgroup
 \patchcmd{\titleblock@produce}
  {\frontmatter@RRAPformat}
  {\frontmatter@RRAPformat{\produce@RRAP{*#1\href{mailto:#2}{#2}}}\frontmatter@RRAPformat}
  {}{}
}%
\makeatother
\begin{document}

\preprint{AIP/123-QED}

\title[]{Revisiting the Siegert relation for the partially coherent regime of nanolasers}
\author{M. Drechsler}
\affiliation{ 
    Institute for Theoretical Physics, University of Bremen, Bremen, Germany
}
\affiliation{
Bremen Center for Computational Materials Science, University of Bremen, Bremen, Germany
}
\author{F. Lohof}%
\affiliation{ 
    Institute for Theoretical Physics, University of Bremen, Bremen, Germany
}%
\affiliation{
Bremen Center for Computational Materials Science, University of Bremen, Bremen, Germany
}
\author{C. Gies}
    \email{gies@itp.uni-bremen.de}
 \homepage{http://www.itp.uni-bremen.de/ag-gies/}
 \affiliation{ 
    Institute for Theoretical Physics, University of Bremen, Bremen, Germany
}%
\affiliation{
Bremen Center for Computational Materials Science, University of Bremen, Bremen, Germany
}

\date{\today}

\begin{abstract}
The Siegert relation connects the first- and second-order coherence properties of light. While strictly valid only in the thermal regime and in the absence of correlations, this relation is routinely extended to the partially coherent regime in the study of high-$\beta$ nanolasers, where it aids in the identification of the lasing threshold. We re-evaluate the use of a generalized Siegert relation in different device regimes. A full two-time quantum-optical theory is derived as a reference for obtaining first- and second-order correlation functions $g^{(1)}(\tau)$ and $g^{(2)}(\tau)$ in the steady state. We find that even in the partially coherent regime the generalized Siegert relation is well suited as an approximation to $g^{(2)}(\tau)$ as long as emitter correlation are negligible, but does not apply well in the quantum regime of few-emitter nanolasers, or to devices featuring sub- and superradiant emission.
\end{abstract}

\maketitle

Photon-correlation functions are a powerful tool in the analysis of light, as they carry signatures of the light's photon statistics, its spatial and temporal coherence, and they contain information on the emission process itself \cite{loudon_quantum_2000}. In particular, the first-order coherence function $g^{(1)}(\tau)$ is related to temporal fluctuations in the electric field and the spectral properties of the light, while the second-order coherence function $g^{(2)}(\tau)$ contains information on intensity fluctuations \cite{glauber_quantum_1963}. These coherence functions are generally an accepted measure of the coherence properties in the emission of nanolaser devices. Nevertheless, it should be said that small lasers may exhibit peculiar properties that arise from many-body effects, correlations, and fluctuations that go beyond what is commonly captured in the first and second moments of the photon statistics \cite{wang_dynamical_2015, PhysRevLett.102.053902, PhysRevApplied.9.064030}.
Analysis of second-order correlations is used in cavity quantum electrodynamics (cQED) \cite{hamsen_two-photon_2017, snijders_observation_2018} and for the study of emission characteristics and noise spectra of LED and laser light sources \cite{paik_interlayer_2019, megyeri_directional_2018, shan_coherent_2021, blazek_unifying_2011, wang_methodological_2021}. 
In the study of high-$\beta$ nanolasers, measurements of first- and second-order coherence are commonly used to reveal the transition from thermal to coherent emission as they carry fingerprints of the corresponding photon statistics \cite{jagsch_quantum_2018, ma_applications_2019, kreinberg_thresholdless_2020,ulrich_photon_2007, takemura_lasing_2019, mork_rate_2018}. 

The Siegert relation \cite{siegert_fluctuations_1943} relates field and intensity correlations for light that exhibit a Gaussian field-amplitude distribution via
\begin{align}
g^{(2)}(\tau) = 1 + |g^{(1)}(\tau)|^2.\label{eq:Siegert_relation}
\end{align}
Its classical derivation assumes a large ensemble of emitters and uncorrelated emission events \cite{voigt_comparison_1994, ferreira_connecting_2020}. However, a quantum mechanical treatment relies on the Wick's theorem for systems in thermal equilibrium and weak interaction between light and matter \cite{fricke_transport_1996} (see supplementary material). The Siegert relation has been employed to analyze experiments in astronomy \cite{foellmi_intensity_2009}, molecular physics \cite{borycki_interferometric_2016}, biology \cite{stetefeld_dynamic_2016}, and quantum optics \cite{asmann_measuring_2010}. In particular, it has become an established tool in the characterization of the laser threshold in nanolasers \cite{ulrich_photon_2007,chow_emission_2014}.

In studies of high-$\beta$ semiconductor nanolasers, the transition from thermal to coherent emission is commonly detected by measurements of the second-order correlation function $g^{(2)}(\tau)$ with a Hanbury-Brown and Twiss (HBT) setup that uses avalanche photo diodes or superconducting single-photon detectors (SSPD) for recording correlated photon-emission events \cite{kreinberg_thresholdless_2020, kreinberg_emission_2017,takemura_lasing_2019,elvira_higher-order_2011,yokoo_subwavelength_2017}. The temporal detector resolution in such experiments is insufficient to resolve photon bunching peaks in $g^{(2)}(\tau)$ due to small coherence times of thermal and only partially coherent light. To address this issue, a generalized Siegert relation (GSR) is assumed to hold in the partially coherent regime, which is characterized by $g^{(2)}(0) < 2$, yet not coherent emission with $g^{(2)}(0) =1$ \cite{blazek_unifying_2011} 
\begin{align}
g^{(2)}(\tau) = 1 + a|g^{(1)}(\tau)|^2.\label{Mod_siegert_relation}
\end{align}
It asserts that intensity correlations captured by $g^{(2)}(\tau)$ are directly related to first-order correlations and decay on a timescale given by the coherence time 
\begin{align}
    \tau_\mathrm{c} = \int_{-\infty}^{\infty} |g^{(1)}(\tau)|^2 \mathrm{d}\tau,\label{eq:tau_c}
\end{align}
which defines the timescale on which $g^{(1)}(\tau)$ vanishes, e.g.~in weakly interacting systems we have $|g^{(1)}(\tau)| = \exp(-|\tau|/\tau_\mathrm{c})$ \cite{loudon_quantum_2000}. The factor $a = g^{(2)}(0) - 1$ accounts for the reduced value of $g^{(2)}(0)$ \cite{blazek_unifying_2011}. Using this relation allows to reconstruct the zero-delay two-photon correlation $g^{(2)}(0)$ from detector-limited HBT measurements. In this letter we investigate under which conditions the GSR is applicable and where it might fail. For this purpose we employ a fully quantum-optical laser model and calculate the two-time coherence functions microscopically, allowing us to explicitly verify the validity of the GSR for different classes of nanolasers. Finally, we discuss the use of the GSR in the analysis of semiconductor nanolaser emission and its implications for the interpretation of experimental results. 

We begin by deriving equations of motion for the two-time correlation functions $g^{(1)}(\tau)$ and $g^{(2)}(\tau)$ directly from a microscopic Hamiltonian. For small systems the full dynamics can be calculated with the Lindblad-von Neumann equation. For many emitters this is not feasible, and we use an approach based on the Heisenberg equation of motion for operators together with the cluster expansion approximation \cite{fricke_transport_1996, leymann_expectation_2014, gies_semiconductor_2007, kira_cluster-expansion_2008}. 
We consider a system of $N$ individual two-level emitters that couple uniformly to a single mode of the electromagnetic field. The light-matter interaction is described in dipole approximation and the corresponding Hamilton operator is given by
\begin{align}
H =  \frac{\varepsilon}{2} \sum_i^N \sigma^{(i)} + \hbar \omega b^\dagger b + \sum_j^N \left(g b^\dagger \sigma^{(j)}_- + g^* b \sigma^{(j)}_+ \right),
\end{align}
where $b$ and $b^\dagger$ are photon annihilation and creation operators, and $\sigma^{(i)}$ denote spin-$\frac{1}{2}$ particle operators acting on the Hilbert space of the two-level system $i$. The $\varepsilon$ denotes the transition energy of a single two-level system, $\omega$ is the frequency of the cavity mode, and $g$ is the light-matter coupling strength. Resonator losses, relaxation, and the pump processes are considered using Lindblad terms with constant rates $\kappa$, $\gamma$ and $P$, respectively. Similar models have been used to describe nanolasers with many \cite{kreinberg_thresholdless_2020}, few \cite{moody_delayed_2018} and even single emitters in a microcavity \cite{Strauf_single2011, reithmaier_strong_2004,gies_strong_2017,laussy_luminescence_2009}.
\begin{figure}
    \begin{center}
    \includegraphics[width=0.483\textwidth]{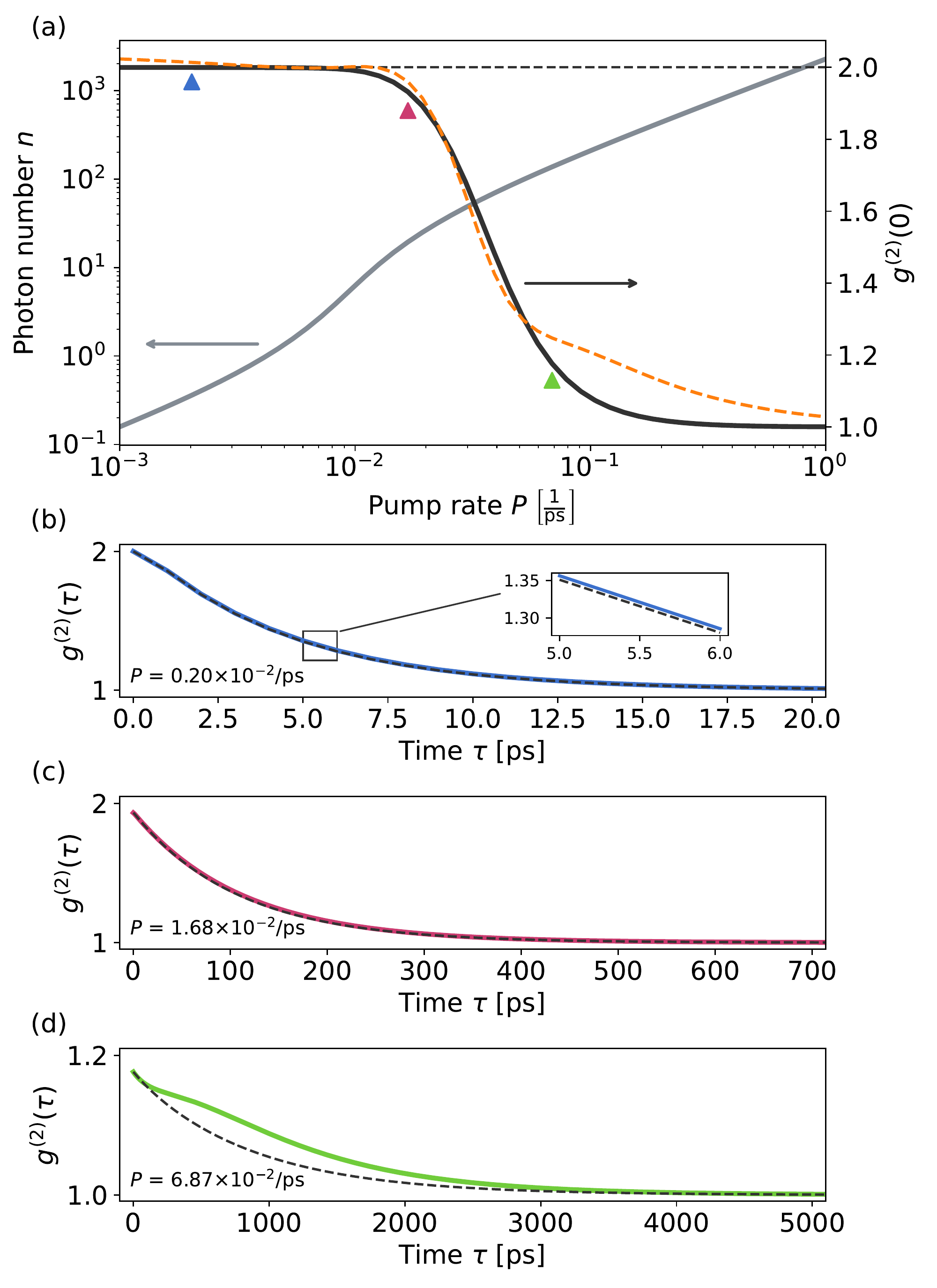}
    \end{center}
    \caption[]{Laser with many emitters subjected to fast dephasing (regime A): $N = 500$, $g = 0.03\,\frac{1}{\mathrm{ps}}$, $\kappa = 0.04\,\frac{1}{\mathrm{ps}}$, $\Gamma = 3\,\frac{1}{\mathrm{ps}}$ and $\beta = 0.1$. (a) Intracavity photon number and second-order correlation $g^{(2)}(0)$ as function of pump power. The orange dashed line shows the reconstructed $g^{(2)}(0)$ calculated according to Eq.~(\ref{g2_entfalten}). Colored markers indicate pump rates at which $g^{(2)}(\tau)$ evaluated in panels below. (b)-(d) Two-time values $g^{(2)}(\tau)$ at fixed pump rates. Dashed lines show predictions by the GSR according to Eq.\ \eqref{Mod_siegert_relation}. The inset in (b) highlights small differences between $g^{(2)}(\tau)$ and Eq.\ \eqref{Mod_siegert_relation} due to residual emitter correlations.}\label{fast_deph}
    \end{figure}
The equation-of-motion approach couples quantities that account for correlations between particles up to arbitrary order in the interaction strength. The arising hierarchy of coupled equations is truncated in the cluster-expansion approximation \cite{kira_quantum_1999} by neglecting correlations of five or more particles, which gives rise to a closed set of equations (see supplementary material). We use the quantum regression theorem \cite{gardiner_quantum_2004} to derive equations of motion for two-time photon-correlation functions using the cluster-expansion. From this we obtain a relation between $g^{(1)}(\tau)$ and $g^{(2)}(\tau)$ that contains the Siegert relation \eqref{eq:Siegert_relation} and a correction term
\begin{align}
g^{(2)}(\tau) = 1 + |g^{(1)}(\tau)|^2 + \delta g^{(2)}(\tau).\label{eq:g2_full}
\end{align}
Eq.~\eqref{eq:g2_full} is a key finding of our work and twofold important. First, we can show that it recovers the Siegert relation \eqref{eq:Siegert_relation} in thermal equilibrium using Wick's theorem for the Matsubara Green function \cite{bruus_many-body_2004}. For a non-interacting system in thermal equilibrium, all pure correlations involving three or more one-particle operators vanish and, in this case, $\delta g^{(2)}(\tau) = 0$. Second, our ansatz enables us to numerically evaluate photon-correlation functions to show that under certain conditions corrections $\delta g^{(2)}(\tau)$ can become sizeable. 

In the following, we consider three specific parameter regimes. The first regime represents a class of nanolasers with many emitters subjected to fast dephasing (regime A), such as quantum-well nanolasers as considered in [\onlinecite{kreinberg_thresholdless_2020}]. Second, we investigate a regime with fewer emitters, in which correlations between individual emitters can form due to their interaction with a common cavity mode (regime B) \cite{jahnke_giant_2016,leymann_sub-_2015}. Finally, we consider the limiting regime of cQED (regime C), in which very few emitters couple strongly to a single laser mode as in [\onlinecite{reithmaier_strong_2004}]. We use our quantum optical model to analyze the behavior of $g^{(1)}(\tau)$ and $\delta g^{(2)}(\tau)$ in all three regimes and subsequently test the validity of the GSR and its applicability for the characterization of nanolasers.

For regime A, Fig.~\ref{fast_deph}(a) displays input-output characteristics and the steady state value of $g^{(2)}(0)$ as a function of pump power. It shows the threshold jump in the photon number (gray) and the transition of the two-photon correlations (black) from the thermal to the coherent regime \cite{chow_emission_2014,jagsch_quantum_2018}. Fig~\ref{fast_deph}(b)--(d) show the correlation function $g^{(2)}(\tau)$ (solid lines) calculated by solving the equations-of-motion for the correlations in Eq.~\eqref{eq:g2_full}. For comparison, we plot the result obtained from the GSR in Eq.~\eqref{Mod_siegert_relation} (dashed lines), which allows comparing the time scales on which first- and second-order correlations decay. We observe that in regime A the GSR gives a good approximation at all pump rates, and in particular in the partially coherent regime for $g^{(2)}(0) > 1.3$. As we observe in the inset of Fig.~\ref{fast_deph}(b), there are small deviations from the GSR even deep in the thermal regime ($g^{(2)}(0) = 2$). This deviation is related to a residual influence of emitter correlations in the system that are treated consistently in our quantum-optical model but are largely suppressed by fast dephasing.

The effect of emitter correlations becomes more evident when the number of emitters is reduced and emitter correlations are not strongly suppressed by dephasing, which is the case in regime B\@. The presence of emitter correlations gives rise to sub- and superradiant coupling, as observed e.g.~in [\onlinecite{kreinberg_emission_2017}] and [\onlinecite{leymann_sub-_2015}].
\begin{figure}
\begin{center}
\includegraphics[width=0.483\textwidth]{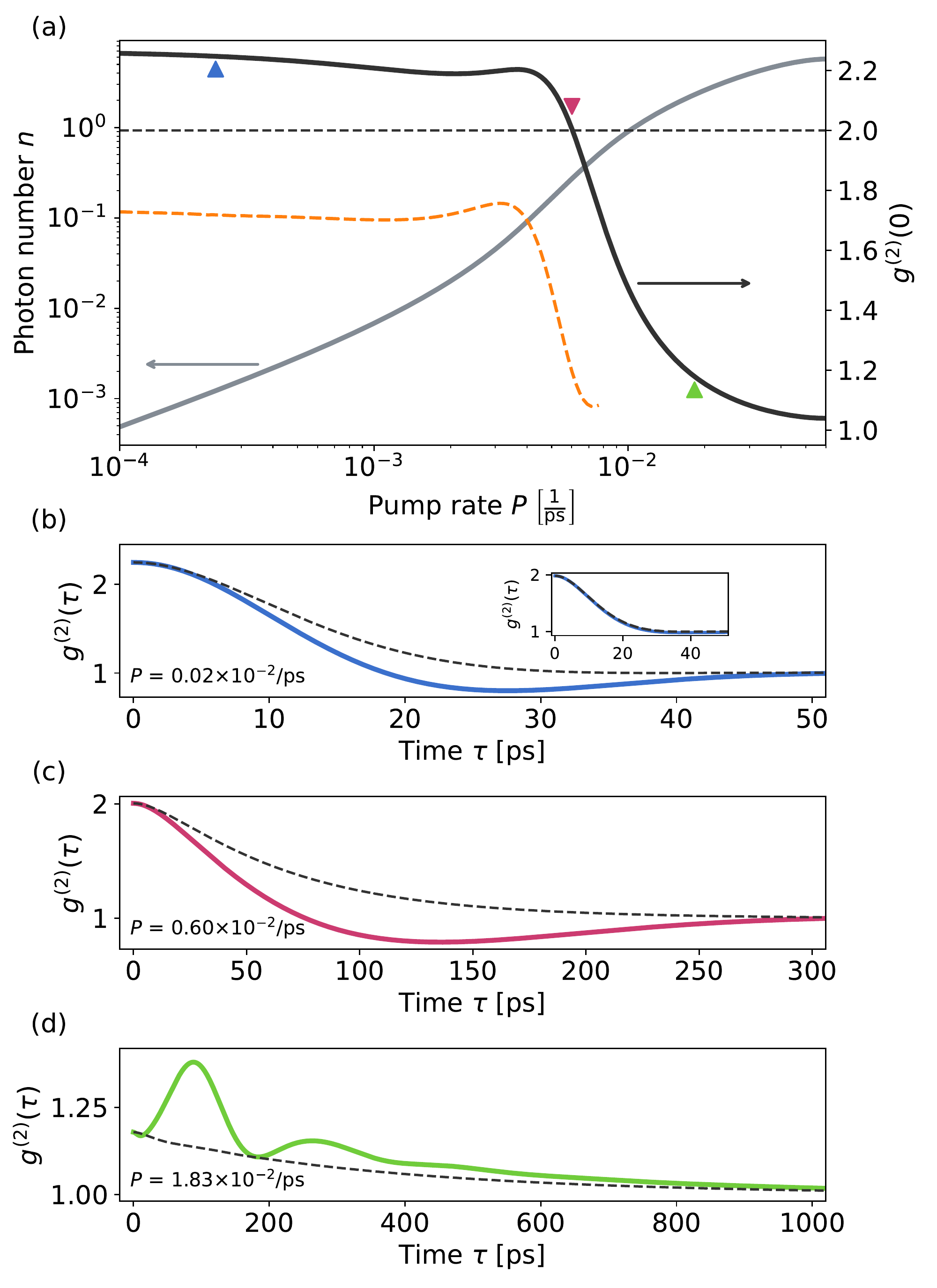}
\end{center}
\caption[]{Laser representing regime B: $N = 100$, $g = 0.008\,\frac{1}{\mathrm{ps}}$, $\kappa = 0.1\,\frac{1}{\mathrm{ps}}$ and $\beta = 0.2$. (a) Intracavity photon number as function of pump power and second-order correlation $g^{(2)}(0)$ showing superthermal emission below threshold due to correlations. The black dashed line marks the value $g^{(2)}(0) = 2$. The orange dashed line shows the reconstructed $g^{(2)}(0)$ calculated according to Eq.~(\ref{g2_entfalten}). Colored markers indicate pump rates at which $g^{(2)}(\tau)$ evaluated in panels below. (b)-(d) Two-time values $g^{(2)}(\tau)$ at fixed pump rates. Dashed lines show predictions by the GSR according to Eq.\ \eqref{Mod_siegert_relation}, which fails to reproduce the antibunching $g^{(2)}(\tau)<1$ at finite delay $\tau$. The inset in (b) shows $g^{(2)}(\tau)$ calculated without emitter correlations which restores agreement with the GSR.}\label{Second_regime}
\end{figure}
In Fig.~\ref{Second_regime} we show results obtained for a prototypical device operating in regime B\@. It indeed exhibits superthermal bunching features at low excitation (Fig.~\ref{Second_regime}(a)), which have been identified as a signature in the subradiant regime \cite{leymann_sub-_2015}. Here, the assumption of individual, uncorrelated emitters is not justified, and consequently we observe deviations from the GSR in panel (b)--(d) at all pump rates. In particular, $g^{(2)}(\tau)$ shows anti-correlated behavior $g^{(2)}(\tau) < 1$ for finite delays $\tau$ that the GSR fails to capture since, by definition, we have $|g^{(1)}(\tau)|^2 > 0$ for all $\tau$. Nevertheless, the GSR correctly quantifies the timescale, on which the second-order correlations decay below the threshold. In Fig.~\ref{Second_regime}(c) we have $g^{(2)}(0) = 2$, which is commonly associated with thermal emission where the Siegert relation is supposed to hold. Still, we see a significant deviation from the results of the full calculation, which is an example of the fact that the zero-delay second-order coherence is an indicator, but not a proof for a particular state of the electromagnetic field. 

To prove that the failure of the GSR is indeed caused by emitter correlations, we take advantage of the possibility to suppress their effect in our theoretical model.
While emitter correlations are inseparable from the rest of the dynamics in a Lindblad-von-Neumann equation, our equation-of-motion approach has the feature that it allows to specifically neglect terms accounting for emitter correlations. In the inset of Fig.~\ref{Second_regime}(b) we show $g^{(2)}(\tau)$ for this case. We observe that the superthermal bunching for $\tau=0$ vanishes, and we recover the physics represented by the Siegert relation. This highlights that the GSR should only be used for the characterization of nanolaser devices when it is clear that emitter-correlation effects related to sub- and superradiance are largely suppressed.

\begin{figure}
    \begin{center}
    \includegraphics[width=0.483\textwidth]{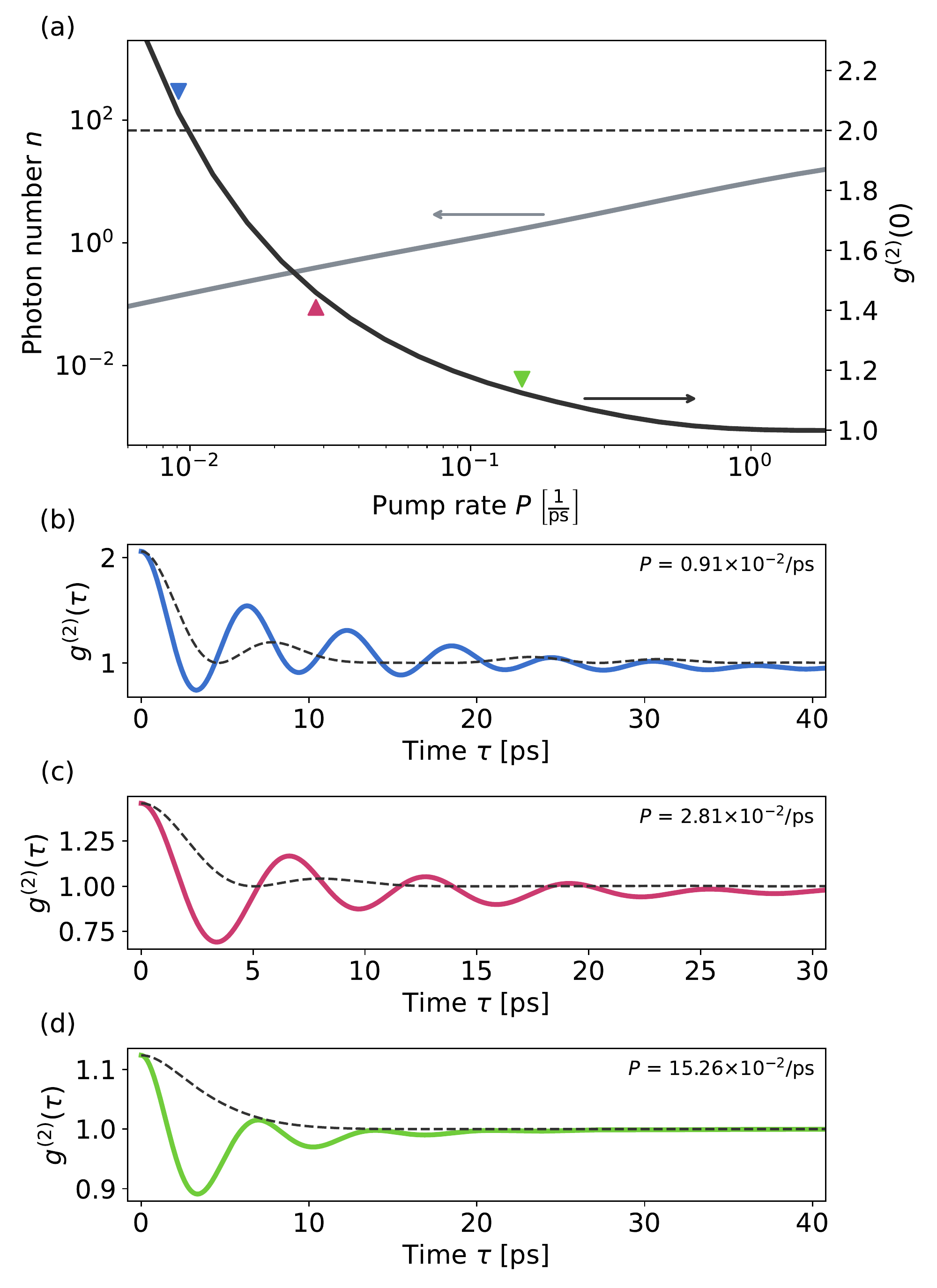}
    \end{center}
    \caption[]{Laser with few emitters operating in regime C: $N = 3$, $g = 0.3\,\frac{1}{\mathrm{ps}}$, $\kappa = 0.07\,\frac{1}{\mathrm{ps}}$, $\beta = 1$. (a) Intracavity photon number and second-order correlation $g^{(2)}(0)$ as function of pump power. Colored markers indicate pump rates at which $g^{(2)}(\tau)$ evaluated in panels below. (b)-(d) Two-time values $g^{(2)}(\tau)$ at fixed pump rates show pronounced oscillations due to strong coupling conditions. Dashed lines show predictions by the GSR according to Eq.\ \eqref{Mod_siegert_relation}, which fail to reproduce the time scales of $g^{(2)}(\tau)$ in this regime.}
    \label{Strong_coupling}
    \end{figure}

Finally, Fig.~\ref{Strong_coupling} shows results for regime C, which is a system of three emitters coupled to a single cavity mode. In this cQED limit, lasing is typically possible in the strong-coupling regime \cite{gies_strong_2017}, and we treat this system exactly by solving the Lindblad-von-Neumann equation, which contains correlations up to arbitrary order in the interaction strength. The parameters are chosen such that the light-matter interaction exceeds all dissipation rates in the system, which is necessary to achieve lasing \cite{gies_strong_2017}. Here we observe even stronger superthermal bunching below threshold than in the previous case. The presence of strong light-matter coupling gives rise to Rabi oscillations, which manifest themselves also in the coherence properties (see Fig.~\ref{Strong_coupling}(b)--(d)). Clearly, the validity of the GSR is no longer given in this regime, and it is no longer justified to use it for the estimation of the time scales, on which the second-order correlations decay.

\begin{figure}
    \begin{center}
    \includegraphics[width=0.483\textwidth]{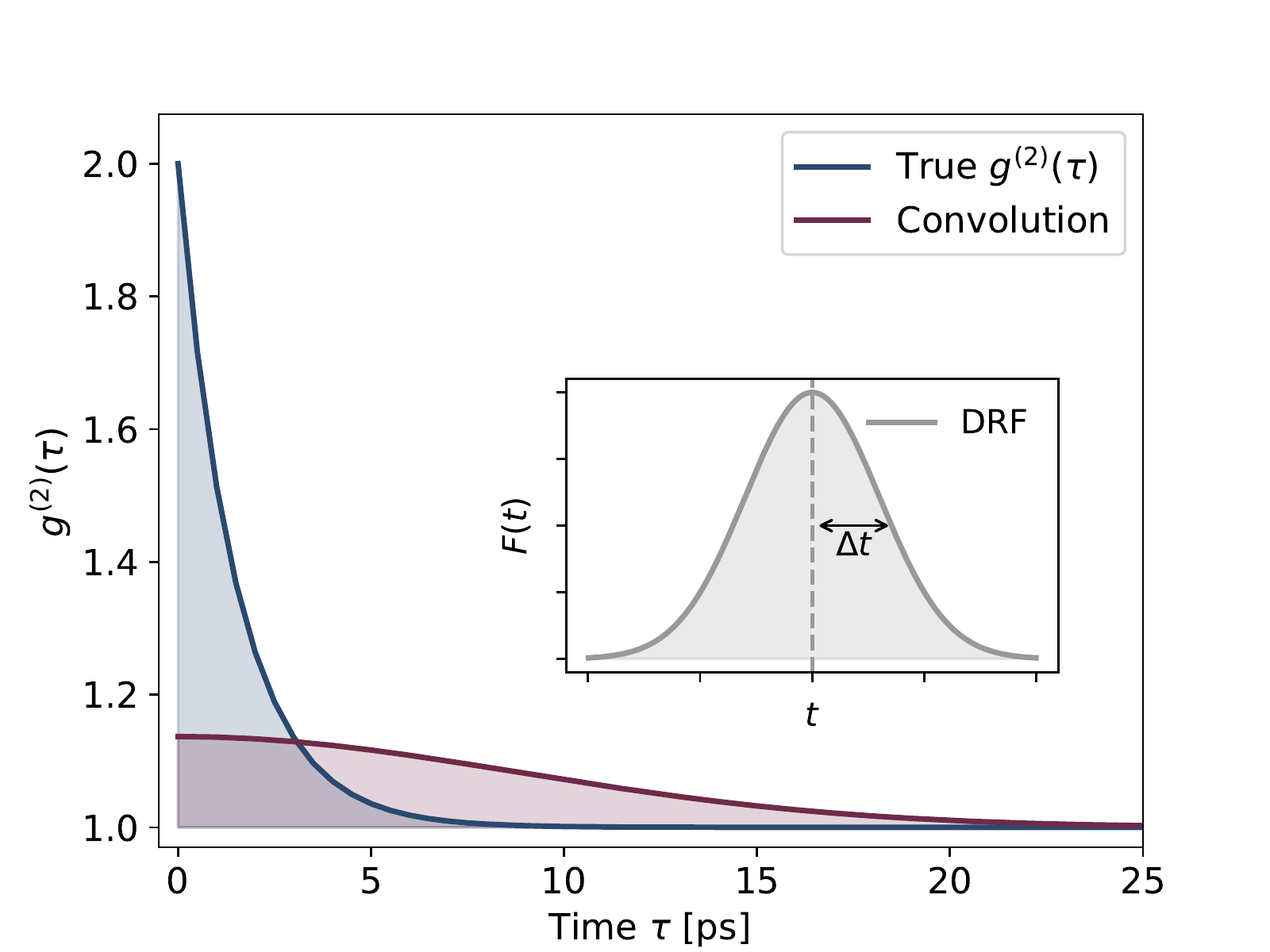}
    \end{center}
    \caption[]{Illustration of the convolution of $g^{(2)}(\tau)$. Here, a thermal $g^{(2)}(\tau)$ ($g^{(2)}(0) = 2$) was convoluted with a DRF modelled by a Gaussian function. The FWHM of the DRF function is given by $\Delta t = 20\,\mathrm{ps}$ and quantifies the timing jitter of the detector.}\label{Convolution_sketch}
    \end{figure}

In the remaining part of this letter, we discuss the applicability of the GSR in experimental characterization of semiconductor nanolasers where the second-order photon-correlation $g^{(2)}(0)$ acts as a fingerprint of the light's photon statistics. It is determined in an HBT setup \cite{jagsch_quantum_2018, hayenga_second-order_2016, kreinberg_thresholdless_2020, ulrich_photon_2007}, in which the used single-photon detectors exhibit a timing jitter, i.e.\ an uncertainty in photon arrival times, limiting the temporal resolution of the measurement \cite{korzh_demonstration_2020}. Together with the small coherence time of the emission for thermal and partially coherent light, this prevents the resolution of photon bunching at zero time delay in this regime. Fig.~\ref{Convolution_sketch} illustrates the averaging procedure caused by the detector's temporal resolution using the example of a thermal $g^{(2)}(\tau)$ (blue line). The timing jitter redistributes the recorded arrival times of photons, which is modelled by a convolution of the correlation function $g^{(2)}(\tau)$
\begin{align}
    \tilde{g}^{(2)}(\tau) = \int_{-\infty}^\infty g^{(2)}(t) F(\tau - t) \mathrm{d}t,\label{g2_convolution}
    \end{align}
with a detector response function (DRF) $F(t)$ shown in the inset of Fig.~\ref{Convolution_sketch}, where the temporal resolution $\Delta t$ of the detector defines the FWHM of the DRF. The result of the convolution corresponds to the \textit{measured} $\tilde{g}^{(2)}(\tau)$ (mauve line in Fig.~\ref{Convolution_sketch}). Clearly, the convoluted $\tilde{g}^{(2)}(0)$ significantly differs from $g^{(2)}(\tau)$, showing strongly reduced signatures of photon bunching $\tilde{g}^{(2)}(0) < 2$ at thermal emission. Crucially, the areas under the curves are invariant under the convolution due to the normalization of the DRF. 
By integrating $g^{(2)}(\tau) - 1$ we find
\begin{align}
&\tau_\mathrm{area} = \int_{-\infty}^\infty (g^{(2)}(\tau) - 1) \mathrm{d}\tau = a \int_{-\infty}^\infty |g^{(1)}(\tau)|^2 \mathrm{d}\tau = a \tau_\mathrm{c},\label{eq:tau_area}
\end{align}
where we used Eq.~\eqref{Mod_siegert_relation} and Eq.~\eqref{eq:tau_c} for the coherence time $\tau_\mathrm{c}$. Thus, assuming that the GSR is valid for all pump rates, it can be used to reconstruct the zero-delay second-order coherence $g^{(2)}(0)$. We can rearrange equation \eqref{eq:tau_area} and insert it into Eq.~\eqref{Mod_siegert_relation} to obtain the \textit{reconstructed} value of $g^{(2)}(0)$:
\begin{align}
g^{(2)}_\mathrm{rec.}(0) = 1 + \frac{\tau_\mathrm{area}}{\tau_\mathrm{c}},\label{g2_entfalten}
\end{align}
i.e., we can determine $g^{(2)}_\mathrm{rec.}(0)$ from quantities that are not subject to the temporal resolution of the detector. As the convolution with the DRF does not change $\tau_\mathrm{area}$, we can integrate $\tilde{g}^{(2)}(\tau)$ directly to obtain $\tau_\mathrm{area}$.

We illustrate the reconstruction process by adding the results for $g^{(2)}_\mathrm{rec.}(0)$ to Figs.~\ref{fast_deph}(a) and \ref{Second_regime}(a)
as orange dashed line. As we have discussed above, the GSR gives a good approximation to the real $g^{(2)}(\tau)$ for the thermal and partially coherent regimes only if emitter correlations are negligible. For regime A, where these conditions are fulfilled the described procedure leads to good agreement with the actual $g^{(2)}(0)$. Below $g^{(2)}(0) = 1.3$ we see a deviation, indicating that the GSR ceases to be valid beyond this point due to oscillations appearing in $g^{(2)}(\tau)$. However, in a real experiment the photon bunching of the thermal regime and most of the transition regime $g^{(2)}(0) < 2$ can be reliably reconstructed with this method. For regime B\@, the reconstructed $g^{(2)}(0)$ severely underestimates the true value which is linked to the anti-correlated behavior of $g^{(2)}(\tau)$ at finite $\tau$. 

In conclusion, we have reviewed the validity of the GSR and its application in the characterization of high-$\beta$ nanolasers with semiconductor based gain media. We have developed a quantum-optical approach to access two-time second-order correlations functions taking into account emitter correlations consistently, giving rise to a correction to the Siegert relation that makes it valid in all emission regimes. In nanolasers with extended inhomogeneous gain media, such as semiconductor quantum wells or ensembles of quantum dots, we expect that fast dephasing suppresses emitter correlations, such that the GSR is largely valid and can be used for reconstructing $g^{(2)}_\mathrm{rec.}(0)$. In the presence of emitter correlations, strong deviations from the exact calculations occur, leading us to conclude that it is imperative to verify the absence of such correlations prior to invoking the GSR. For future studies our equation of motion technique can be extended to access $g^{(2)}(\tau)$ for a range of different nanolaser devices, such that relying on the GSR becomes unnecessary. 
\\
See the supplementary material for the details regarding the quantum mechanical treatment of the Siegert relation and the derivation of the quantum-optical laser model.
\\
F.L.~acknowledges funding by the central research development fund (CRDF) of the University of Bremen.

\section*{AUTHOR DECLARATIONS}
\subsection*{Conflict of Interest}
The authors have no conflicts to disclose.
\section*{DATA AVAILABILITY}
The data that support the findings of this study are available from the corresponding authors upon reasonable request.
\bibliography{Drechsler_et_al_final}

\end{document}